# Longitudinal tri-foci Metalens empowered multiple-magnification and diffraction-limited microscope

*Chuang Sun,[1] Zixuan Wang,[1] Kian Shen Kiang,[1] Jun-Yu Ou,[2*] and Jize Yan,[1*]*

[1] University of Southampton, School of Electronics and Computer Science, Southampton SO17 1BJ, UK

[2] University of Southampton, Department of Physics and Astronomy, Southampton SO17 1BJ, UK

**ABSTRACT** Dielectric metalens has emerged as an attractive device for advanced imaging system because of its powerful manipulation ability of light beam, small volume, and light weight. However, the applications of silicon nitride ($Si_3N_4$) metalens are limited by the low refraction index of $Si_3N_4$, and multi-foci metalens has not been realized based on a $Si_3N_4$ metalens. Here, we deeply explore the working mechanism of a truncated waveguide meta-atom and obtain a $Si_3N_4$ metalens with longitudinal three diffraction-limited focal points. By utilizing the metalens sample as a condenser lens, a commercial microscope can obtain three magnifications based on a single objective lens. Finally, an infinity-corrected microscope with three high magnifications (9.5X, 10X, and 29X) and diffraction-limited resolution is integrated into centimeter-dimension for the first time by using the tri-foci metalens sample as an objective lens. This research would boost the scaling up of metalens microscope as well as the multifunctional application of $Si_3N_4$ metalens.

Since Capasso's group reported the abrupt phase change over the wavelength-scale and the generalized Snell law in the year of 2011 [1], the research on metasurface have experienced explosive growth around the world because the metasurface's powerful engineering ability of electromagnetic wave's amplitude [2, 3], phase [4, 5], polarization [6, 7], and even frequency [8, 9]. Comparing with plasmonic metasurface, all-dielectric metasurface is attracting intensive attention because it can avoid ohmic losses and realize higher working efficiency [10]. The working mechanism of an all-dielectric metasurface can be based on resonance effects (e.g., Mie-resonance, Fano resonance, bound states in the continuum, et al.) or non-resonance effects (e.g., propagation phase, geometry phase, and their combination) [10-13]. The imaging application of resonance phase metalens (i.e., metasurface with a focusing phase profile) would be restricted by its narrow working bandwidth and the strong resonance coupling between adjacent meta-atoms [14]. In contrast, metalens based on non-resonance phase can work in a wide spectrum and work for tightly focusing with a high numerical aperture (NA). For example, the metalens based on truncated waveguide meta-atom has been adopted for achieving high-efficiency, high-resolution, and broadband imaging in ultraviolet, visible, and infrared spectrum [15-19].

While the geometry phase is achieved by using anisotropic meta-atoms, the propagation phase introduced by a meta-atom generally requires a wavelength-scale height and a material having high refraction index ($n$) comparing with the environment ($n_e$) and low absorption index [11, 12]. Therefore, silicon is generally used in near infrared and short-wavelength mid-infrared spectrum, the $TiO_2$, GaN, and $Si_3N_4$ are widely used in visible spectrum. Comparing with CMOS-compatible fabrication process of $Si_3N_4$ metalens, the fabrication of $TiO_2$ and GaN metalens is not compatible with CMOS process and require a complex process of low-temperature atomic layer deposition and high aspect ratio two-step dry etching. However, the refraction index of $Si_3N_4$ is around 2 and smaller than that (~2.4) of $TiO_2$ and GaN, which would lead to a high aspect ratio for achieving $2\pi$ phase coverage and increase the fabrication difficulty [11]. As a result, the multi-functional applications of a $Si_3N_4$ metalens is restricted. Because the low-cost and CMOS-compatible fabrication process are two key factors of boosting the industrial applications of dielectric metalens, it is meaningful to explore the potential of a $Si_3N_4$ metalens for multi-functional applications. For example, a single metalens can integrate multiple imaging systems into one [21, 22].

Different from a conventional lens, a metalens can be engineered to possess multiple focal points via spatial multiplexing or wavefront shaping technique, which makes it possible to realize a multiple imaging system via single metalens [22-25]. However, the multiple-foci metalens has not been used to build up an integrated microscope system with multiple magnifications and diffraction-limited resolution. In addition, there is no report on $Si_3N_4$ multi-foci metalens.

In a conventional microscope, only one magnification and resolution can be achieved using a single objective lens [26]. Multiple objective lenses are necessary for realizing multiple magnifications and resolutions, and a nosepiece with high accuracy is required for replacing the objective lens [the section marked by green rectangle in Fig. 1(a)]. As a result, the conventional microscope system is complicated and bulky. To integrate three objective lenses into one element and eliminate the nosepiece, a 1.2mm $Si_3N_4$ metalens with longitudinal three focal points (i.e., tri-foci metalens, t-MS) is obtained in this paper [Fig. 1(b)]. By introducing the t-MS into a commercial microscope [Fig. 1(a)], three magnifications can be obtained without replacing objective lens. In last of this paper, the whole microscope system [the section marked by purple rectangle in Fig. 1(a)] including three objective lenses is miniaturized into two metalens samples [Fig. 1(e)], and two samples are compactly integrated by a 3D printed holder [Figs. 1(c) and 1(d)]. In the miniaturized microscope system, the t-MS is utilized as an objective lens with three focal points, and a single-foci metalens (*q*-MS) with a quadratic phase profile serves as a tube lens [Fig. 1(e)].

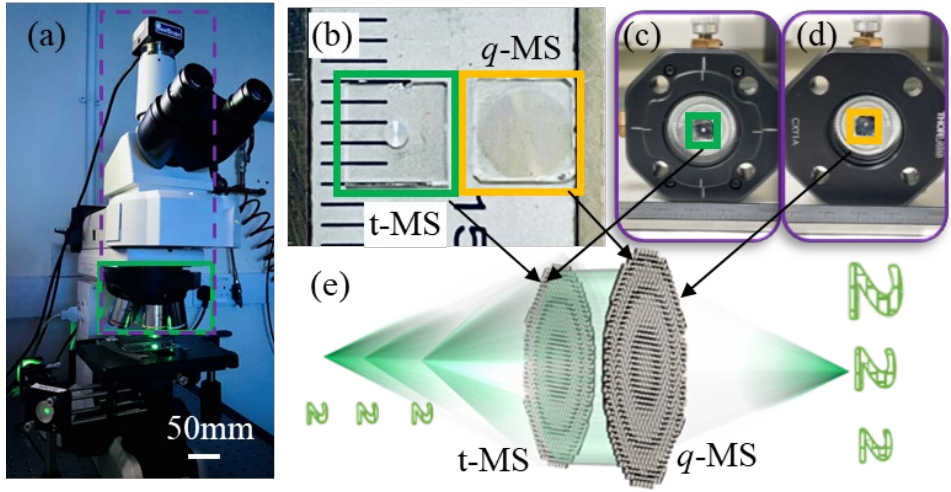

**Figure 1.** Conceptual figure of integrated microscope with three magnifications. (a) Conventional infinity-corrected microscope; (b) optical images of two metalens samples; (c) and (d) are the front (t-MS) and rear (*q*-MS) surface of the integrated microscope, respectively; (e) Optical configuration of integrated microscope. t-MS: tri-foci metalens; *q*-MS: metalens with quadratic phase profile.

To achieve the t-MS based on $Si_3N_4$ with a refraction index of 2, a general working mechanism of an anisotropic truncated waveguide meta-atom [Fig. 2(a)] is investigated (see section 1 in supporting material). It is figured out that the ongoing light $|E_o\rangle$ from the meta-atom could be imposed three phase profiles $\Phi_1(x,y) = 2\theta(x,y) + \varphi_p(x,y)$, $\Phi_2(x,y) = -2\theta(x,y) + \varphi_p(x,y)$, and $2\Phi_3(x,y) = 2\varphi_p(x,y) = \Phi_1(x,y) + \Phi_2(x,y)$, when the phase delay $\Phi_d$ between X- and Y-polarization light is not equal to zero and $\pi$. $\varphi_p(x,y)$ is the distribution of propagation phase $\varphi_p = (\varphi_x + \varphi_y)/2$, and $2\theta(x, y)$ is the rotation distribution of the meta-atoms. In principle, once a meta-atoms library is built, we can design the metalens by searching the library. In the library, each meta-atom should possess a fixed phase delay ranging from 0 to $\pi$, and the propagation phase $\varphi_p$ of all meta-atoms can achieve $2\pi$ phase coverage.

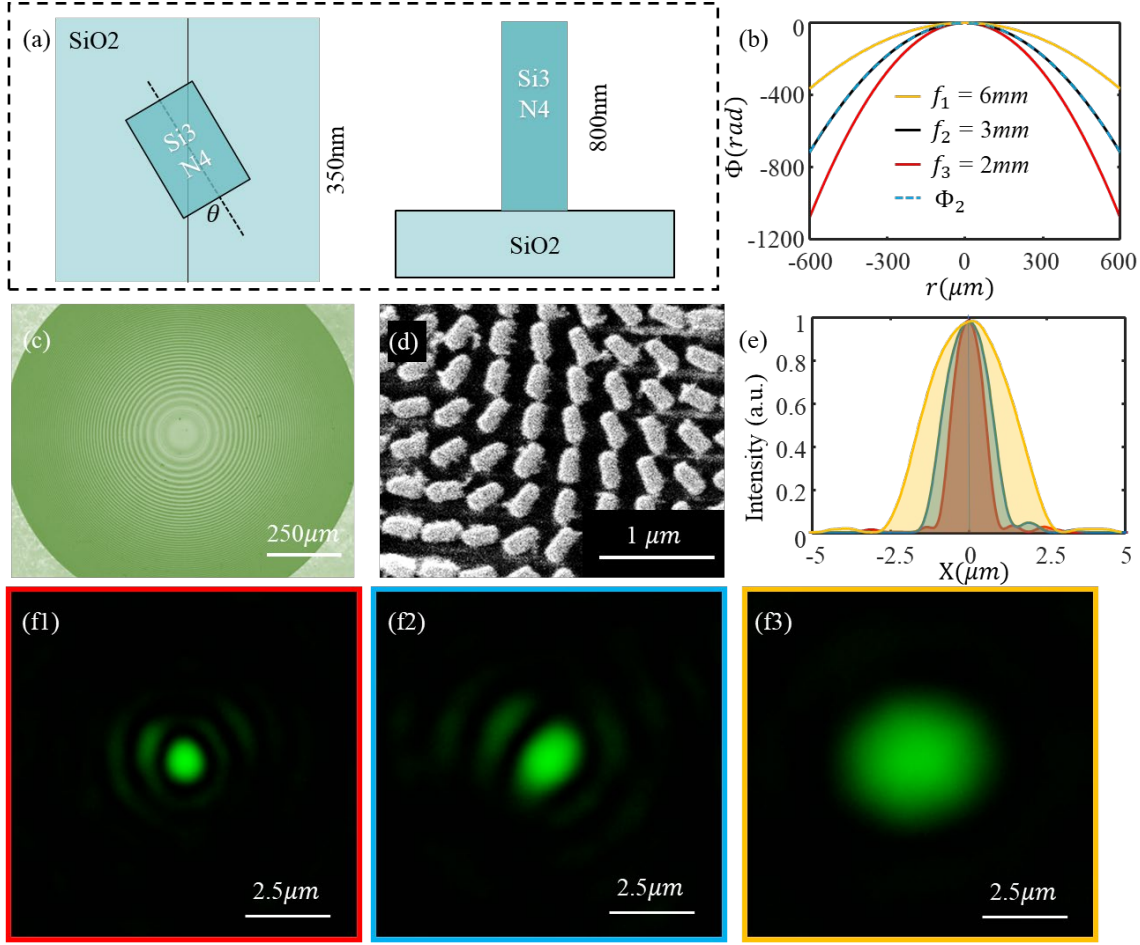

**Figure 2.** Design and optical characterization of the metalens sample. (a) nanofin structure; (b) three hyperbolic phase profiles for the three focal points; (c) optical image of the metalens with a diameter of 1.2mm; (d) metalens sample's SEM image; (e) the FWHMs of three focal points; (f1) - (f3) are the three focal points at focal length $f = 1.9mm, 3mm$, and $6mm$, respectively.

In general, the application of Si3N4 is limited by that the low refraction index makes that it is impossible to realize the classical spin-multiplexing based on library where each meta-atom need possess a phase delay of $\pi$ and all meta-atoms' propagation phase $\varphi_x$ should achieve $2\pi$ phase coverage. According to the analysis in supporting material in section 1, the strict requirement of a phase delay of $\pi$ can be released to an arbitrary constant value between 0 and $\pi$. Therefore, to impose three focusing phase profiles and obtain three longitudinal focal points, we build up a meta-atoms library where each meta-atom possesses a phase delay of $5\pi/12$ and all meta-atoms' propagation phase $\varphi_p$ can achieve $2\pi$ phase coverage (see Fig. S2 in supporting material).

In designing the t-MS, the phase profile $\Phi_1(x, y)$ is the hyperbolic phase profile [Eq. (1)] with a focal length $f_1$ of 6mm, the phase profile $\Phi_3(x, y)$ is the hyperbolic phase profile with a focal length $f_3$ of 2mm [17]. Then, we obtained the $\Phi_2(x, y) = 2\Phi_3(x, y) - \Phi_1(x, y)$. According to Eq. 2, the $\Phi_2(x, y)$ can be seen as a hyperbolic phase with a focal length $F = \frac{2f_1 f_3}{f_1 + f_3}$ =3mm. The three phase profiles are plotted in Fig. 2(b) in yellow, blue, and red colors, respectively. In addition, from the black curve and blue dashed curve, the $\Phi_2(x, y)$ meets well with the hyperbolic phase profile $\Psi(f_2 = 3, x, y)$.

$$\Psi(f, x, y) = -\frac{2\pi}{\lambda}(\sqrt{f^2 + x^2 + y^2} - f) \tag{1}$$

$$2\Psi\left(\frac{2f_1 f_3}{f_3 + f_3}, x, y\right) \approx \Psi(f_1, x, y) + \Psi(f_3, x, y) \tag{2}$$

The t-MS sample is fabricated following a CMOS-compatible flow (see supporting material Figs. S3 and S4). Figs. 2(c) and 2(d) shows the optical image and SEM image of the fabricated metalens sample. Then, the t-MS sample is characterized by a home-built microscope (supporting material Figs. S5). We observed the first focal point [Fig. 2(f1)] with moving the imaging part of the microscope 1.9mm away from the surface [Fig. 2(c)] of the t-MS sample. It means the first measured focal point has a focal length of 1.9mm. With moving microscope's imaging part 3mm and 6mm away from the sample surface, we respectively observed the second [Fig. 2(f2)] and third focal point [Fig. 2(f3)]. Therefore, we obtained three longitudinal focal points with expected focal length. The NAs of three focal points can be calculated as 0.3 ($f = 1.9mm$), 0.2 ($f = 3mm$), and 0.1 ($f = 6mm$), respectively.

To evaluate the focusing performance of the three focal points, we calculated their full-width-of-half-maximum (FWHM) [Fig. 2(e)]. The FWHMs of three focal points are $1.18\mu m$, $1.79\mu m$, and $3.51\mu m$, respectively. As the diffraction limitation ($0.61\lambda/NA$) can be theoretically calculated upon the working length $\lambda = 520nm$ and NA, the diffraction limitations of each focal points can be obtained as $1.02\mu m$ ($f = 1.9mm$, NA = 0.3), $1.56\mu m$ ($f = 3mm$, NA =0.2), and $3.12\mu m$ ($f = 6mm$, NA = 0.3). Comparing the FWHMs and theoretical diffraction limitations of each focal points, it can be concluded that we obtained three near-diffraction-limitation longitudinal focal points via a $Si_3N_4$ metalens.

The t-MS sample is firstly used as a condenser in a commercial microscope [Fig. 1(a)]. As shown in Figs. 3(a) and 3(b), a positive USAF 1951 resolution target (RT) is placed in the collimated illumination light path, and it is illuminated by a lamp with a wide emission spectrum. Then, through the imaging of t-MS sample, three images [Fig. 3(a)] corresponding to the three focal points can be obtained. Then, the three images are collected and magnified by the commercial microscope with an objective lens (10X, NA = 0.3). Because the chromatic aberration is not corrected in designing, the RT cannot be imaged to a same plane for different wavelength at each focal point. Therefore, a tunable bandpass filter (Thorlabs: KURIOS-WB1/M) is placed in front of the camera to capture the image of each single wavelength. Note that the light passing through the filter still has a bandwidth of 35nm. Therefore, the image captured by the camera is not exactly for a single wavelength.

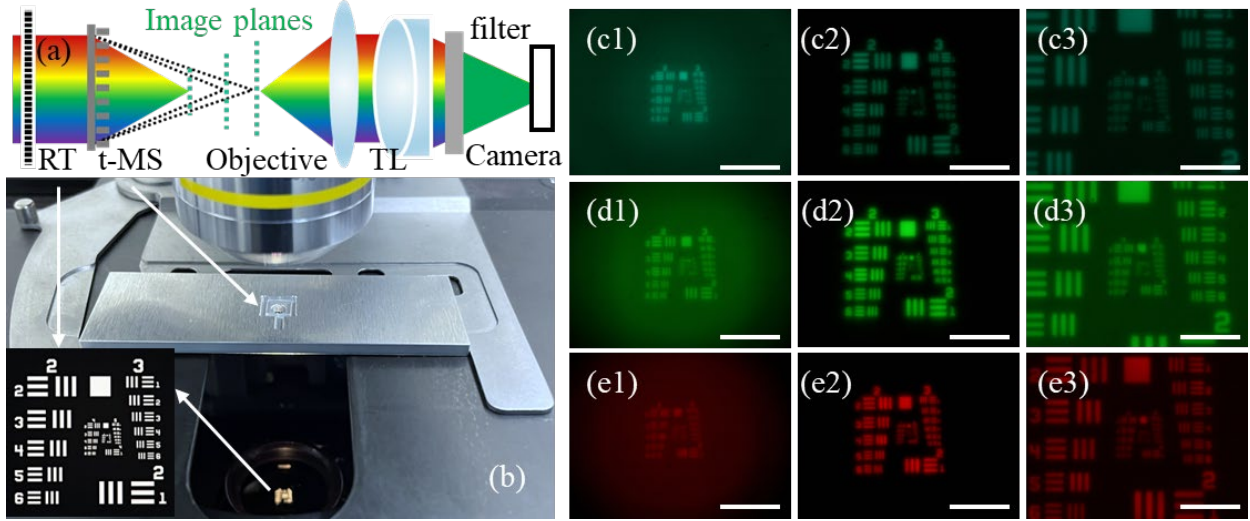

**Figure 3.** Singlet imaging experiment in visible spectrum. (a) Optical configuration. RT: resolution target [USAF 1951, see the insert in (b)]; TL: tube lens; (b) Experiment setup; (c) – (e) are the imaging results (the 1$^{st}$ column is obtained at $f = 1.9$mm, the 2$^{nd}$ column at $f = 3$mm, the 3$^{rd}$ column at $f = 6$mm) for the wavelength of 488nm, 520nm, and 650nm, respectively. Insert figure in (b) is the ground truth of the RT. Scale bars: 1mm.

By setting the central wavelength of the filter at 488nm, 520nm, and 650nm, we obtained 9 images as shown in Figs. 3(c)-3(e). To compare the magnification's difference of each focal point as well as the central wavelength, we marked the scale bars based on the size (i.e., pixel number × pixel size) of image in camera coordinator. Therefore, the larger field of view (FOV) from the images illustrates a smaller magnification. The images on Fig. 3(c1), Fig. 3(c2), and Fig. 3(c3) are respectively obtained at the 1$^{st}$ ($f$ = 1.9mm), 2$^{nd}$ ($f$ = 3mm), 3$^{rd}$ ($f$ = 6mm) focal point. We can see that, with moving the objective lens from the 1$^{st}$ focal point to the 3$^{rd}$ one, the imaging FOV goes down. Because the singlet imaging magnification goes up with the increasement of focal length for a fixed object distance between the RT and t-MS sample.

Comparing Figs. 3(d) and 3(e), it is observed that the imaging FOV goes up with the working wavelength is tuned from 520nm to 650nm, which indicates that the magnifications corresponding to each focal point goes down with increasing the working wavelength. Furtherly, this phenomenon arising from the dispersion relation (i.e., the focal length goes down with increasing the working wavelength) of a metalens. Therefore, we demonstrated the t-MS sample work well for a wide spectrum (488nm – 650nm), while it cannot achieve broadband achromatic imaging. More importantly, we show that a commercial Nikon microscope could realize three magnifications without replacing objective lens by using the t-MS sample as a condenser.

In following, the t-MS sample is adopted as an objective lens to build an integrated and miniaturized microscope with three magnifications. To try miniaturizing the whole microscope to be centimeter level and reduce its weight, we designed and fabricated another metalens sample which is used a tube lens as shown in Figs. 1 and 4(a). To realize large FOV imaging with low distortion, a quadratic phase profile [Eq. (3)] is adopted in designing the second metalens sample (Supporting material section 5) [25]. Therefore, the second metalens sample is called $q$-MS in following. In addition, the $q$-MS has a diameter of 4.5mm to increase the image edge's brightness. As shown in Fig. 4(b), to effectively collecting the light scattered from the off-axis object point, the $q$-MS must have a larger diameter than the $t$-MS [26]. Otherwise, the edge light energy would be lost in the imaging system. The focal length of the $q$-MS is 45mm for achieving high magnifications.

$$\Psi_q(f,x,y) = -\frac{\pi(x^2+y^2)}{\lambda f} \quad (3)$$

As shown in Figs. 1(c) and 1(d), the two metalens samples are integrated together by a 3D printed holder. As the holder is printed using transparent material, the light beam (i.e., stray light) out of the t-MS region would go into the $q$-MS and imaged to the camera as well, which would degrade the integrated microscope's imaging quality. Therefore, the central region of the holder is stained to be black to block the stray light. Based on the holder, the two metalens samples have a separation of 5mm.

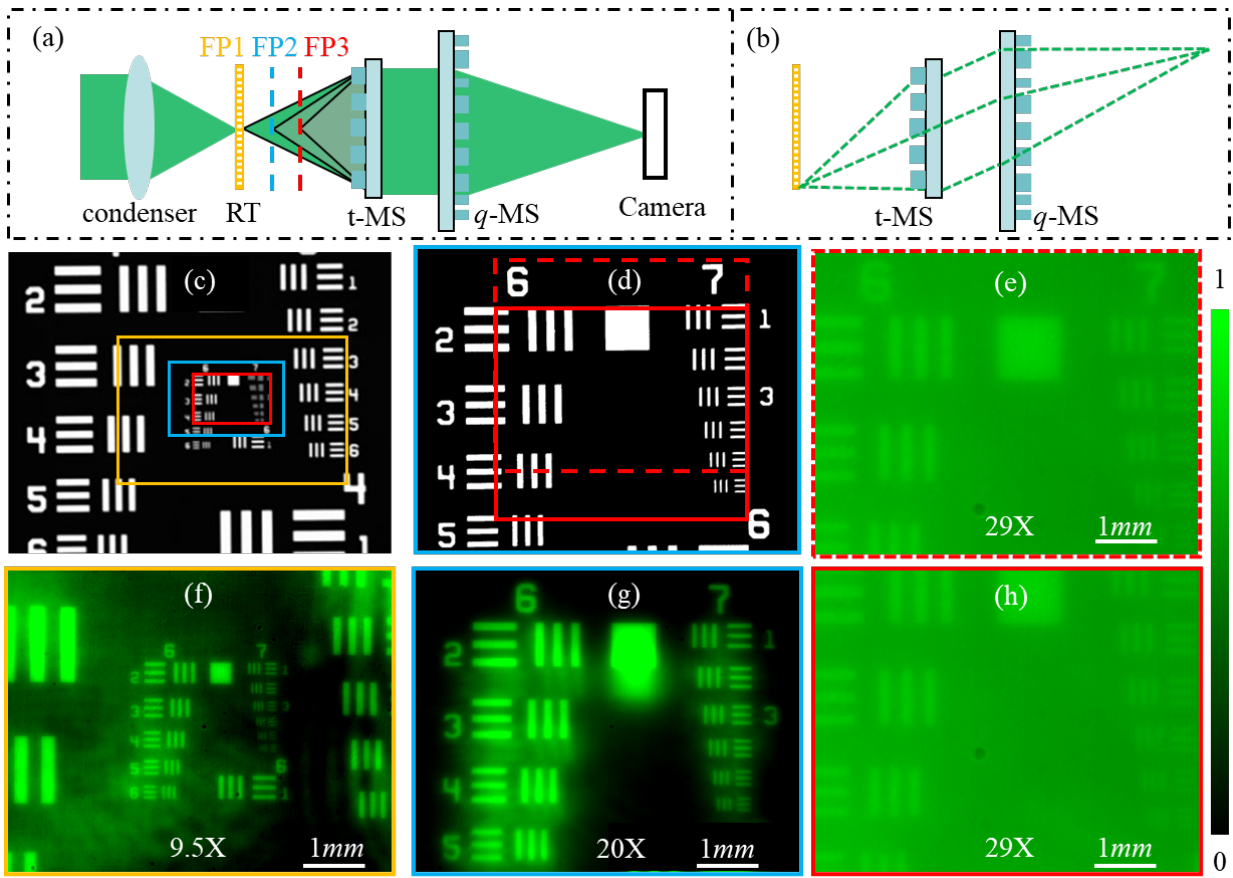

**Figure 4.** Integrated microscope experiment. (a) optical configuration; (b) Optical path of wide FOV imaging; (c) and (d) indicate the imaged FOV for each focal plane; (f) – (h) are the captured images when the RT is respectively placed at $f = 6mm$, 3mm, and 1.9mm; (e) is the second image captured at $f = 1.9mm$.

As shown in Fig. 4(a), to realize intensive illumination, the collimated light from a 520nm LED is focused onto the RT via a condenser lens. The condenser lens and RT are mounted together so that they can be moved from focal plane 1 (FP1) to focal plane 3 (FP3). As the integrated microscope is an infinity-corrected configuration, once the RT is placed in the three focal planes, it can be imaged to the focal plane of the $q$-MS. Therefore, a camera is placed at the $q$-MS focal plane to obtain the images.

Figs. 4(f) – 4(h) illustrate the captured whole images when the RT is placed at $f = $ 6mm [focal plane 1 marked in orange in Fig. 4(a)], 3mm [focal plane 2 marked in blue in Fig. 4(b)], and 1.9mm [focal plane 3 marked in red in Fig. 4(b)], respectively, and the measured FOVs are limited by the camera's sensor size (6.4mm × 5.12mm). The imaged regions as well as the FOVs of each focal plane are marked in orange, blue, and red on a ground truth image [Fig. 4(c)]. As the red FOV ($674\mu m \times 539\ \mu m$) is smaller than the blue one ($320\mu m \times 256\ \mu m$) which is smaller than the orange one ($221\mu m \times 177\mu m$), the FOV goes down with the RT being moved from $f = $ 6mm [Fig. 4(f) orange border] to $f = $ 1.9mm [Fig. 4(h) red border]. Furthermore, it means that the magnifications go up with moving the RT from $f = $ 6mm focal plane [Fig. 4(f)] to $f = $ 1.9mm focal plane [Fig. 4(h)]. Based on quantitative calculation, the magnifications for Figs. 4(f), 4(g), and 4(h) are respectively 9.5X, 20X, and 29X, which are around 1.3 times of the expected magnifications of 7.5X, 15X, and 22.5X. The expected magnifications are evaluated by dividing the focal length 45mm of the $q$-MS by the three focal length 6mm, 3mm, and 1.9mm. We consider that the slightly defocus of the camera contributes to the larger magnifications. When the RT is placed in the focal plane 3, as indicated by the Fig. 4(d), we displaced the imaging region from the red solid rectangle to the red dashed rectangle for obtaining solid evidence of the imaging position (i.e., the group and element number).

Except for the magnification, the imaging resolution of integrated microscope is evaluated as well. From the Fig. 4(f), we can see the element 5 in group 7 with clear contrast, which indicates that the integrated microscope can realize a diffraction-limited resolution of 203.2lp/mm in the whole FOV at the focal plane 1 (focal length = 6mm, NA = 0.1). As the other two focal points have diffraction-limited FWHM as well, we claim that the integrated microscope can achieve diffraction-limited resolutions (322.5lp/mm and 456lp/mm) at the focal planes 2 and 3 as well, while the imaging contrast of the focal plane 3 is lower.

In conclusion, a $Si_3N_4$ t-MS sample with a diameter of 1.2mm is obtained at first. Three diffraction-limited focal points are achieved in experiment. We firstly demonstrated that a commercial Nikon microscope could realize three magnifications with a fixed objective lens by using the t-MS sample for singlet imaging. The broadband imaging performance of the t-MS sample was investigated. Finally, we integrated an infinity-corrected microscope with three high magnifications of 9.5X, 20X, and 29X to a dimension of centimeter level by using the t-MS sample as an objective lens. We conceive the experimentally demonstrated centimeter-scale microscope would boost the metalens microscope's applications in both scientific research and industry fields. In addition, the reported $Si_3N_4$ metalens would pave the way of the multi-functional applications of $Si_3N_4$ metalens.

**Corresponding Author**

**\***Jun-Yu Ou**,** E-mail: bruce.ou@soton.ac.uk

**\*** Jize Yan**,** E-mail: J.Yan@soton.ac.uk

**Author Contributions**

The manuscript was written by Chuang Sun and modified by Jun-Yu Ou, and Jize Yan. The design and fabrication of metalens, characterization and imaging experiments are operated by Chuang Sun. Kian Shen Kiang contributes to the fabrication of the metalens samples. All authors have given approval to the final version of the manuscript.

**Funding Sources**

This work is supported by the UK funding agency EPSRC under grants EP/V000624/1, EP/X03495X/1, EP/T02643X/1 and the Royal Society RG\R2\232531

ACKNOWLEDGMENT

PhD studentship from the Chinese Scholarship Council is acknowledged.